\documentclass[prb,preprint,superscriptaddress,groupedaddress,showpacs,preprintnumbers,amsmath,amssymb]{revtex4}

\usepackage{graphicx}
\usepackage{dcolumn}
\usepackage{bm}

\begin{document}


\title{Optical matrix elements in tight-binding models with overlap}


\author{Titus Sandu}
 \email{tsandu@asu.edu}
\affiliation{Department of Chemical and Materials Engineering,
Arizona State University, Tempe, AZ, 85287}%
  



\date{\today}

\begin{abstract}
We investigate the effect of orbital overlap on optical matrix elements in empirical 
tight-binding models. Empirical tight-binding models assume an orthogonal basis of (atomiclike) 
states and a diagonal coordinate operator which neglects the intra-atomic part. It is shown
that, starting with an atomic basis which is not orthogonal, the orthogonalization process 
induces intra-atomic matrix elements of the coordinate operator and extends the range 
of the effective Hamiltonian. We analyze simple tight-binding models and show that non-orthogonality 
plays an important role in optical matrix elements. In addition, the procedure gives formal justification 
to the nearest-neighbor 
spin-orbit interaction introduced by Boykin [Phys. Rev \textbf{B} 57, 1620 (1998)] in order 
to describe the Dresselahaus term which is neglected in empirical tight-binding models. 
\end{abstract}


\maketitle


\section{Introduction}

The tight-binding (TB) approach to electronic structure is one of the most used 
methods in solid state systems\cite{Turchi98}. The empirical tight-binding (ETB) method, 
which dates back to the work of Slater and Koster\cite{Slater&Koster54}, assumes 
mostly two-center approximation and the matrix elements of 
the Hamiltonian between orthogonal and atom-centered orbitals \cite{Lowdin50} are 
treated as parameters fitted to experiment or first-principles calculations. 
ETB is widely employed to the description of electronic structure of 
complex systems \cite{Delerue01} like interfaces and defects in crystals, amorphous materials, 
nanoclusters, and quantum dots because it is computationally 
efficient (up to 3 order of magnitude faster than the \textit{ab intio} density 
functional methods) and provides physically transparent results. Many calculations 
consider just the nearest-neighbor Hamiltonian with fewer parameters but with additional 
orbitals introduced \cite{Jancu98}. To consider a higher accuracy, the range of 
the Hamiltonian is extended to few nearest-neighbor shells (up to the first 
three shells) and, therefore more fitting parameters \cite{Papaconstantopoulos03}. 
However the use of a non-orthogonal formalism might scale back 
the range of the Hamiltonian having the additional fitting from the overlap matrix. 
In some instances \cite{Boykin02}, when strain is present, 
non-orthogonality is invoked implicitly to accommodate the changes of the 
on-site energies due to local displacements in addition to the well known 
scaling of the transfer integrals \cite{Harrison99}. A non-orthogonal formalism has 
also a less obvious advantage. Because they have a longer range than the atomic orbitals, 
the orthogonalized orbitals samples the local environment, making them better 
suited for transferability to complex systems \cite{Nguyen-Manh00}.

Calculation of optical spectra in the ETB formalism requires the 
knowledge of additional parameters: the momentum or velocity matrix elements 
between initial and final states. In the early work, momentum matrix 
elements were considered as extra parameters fitted to the experimental or 
first-principles calculated dielectric function. However, ETB has been 
extended to include the interaction with electromagnetic fields \cite{Graf95} 
by making the substitution $p = \left( {\raise0.7ex\hbox{${m_0 }$} \!\mathord{\left/ 
{\vphantom {{m_0 } \hbar }}\right.\kern-\nulldelimiterspace}\!\lower0.7ex\hbox{$\hbar $}} 
\right)\nabla _k H$, such that dielectric function and other optical 
properties can be calculated without additional parameters. The scheme is 
based on the Peierls substitution of Hamiltonian matrix elements \cite{Peierls33} allowing us
to calculate directly the momentum or velocity matrix elements. In Ref. \onlinecite{Cruz99} 
and \onlinecite{Pedersen01} it is shown that the substitution $p = \left( 
{\raise0.7ex\hbox{${m_0 }$} \!\mathord{\left/ {\vphantom {{m_0 } \hbar 
}}\right.\kern-\nulldelimiterspace}\!\lower0.7ex\hbox{$\hbar $}} 
\right)\nabla _k H$ leads to the neglect of the intra-atomic momentum matrix 
elements or, equivalently, the coordinate operator 
is diagonal in the subsequent basis as we will indicate below. However, the Peierls-tight-binding 
(i.e. zero intra-atomic position parameters) has been successfully used in 
Ref. \onlinecite{Jancu04}. Pedersen \textit{et al.} 
\cite{Pedersen01} introduced an additional momentum matrix element to accommodate the 
intra-atomic transitions. In contrast, Boykin and Vogl \cite{BoykinVogl02} showed that 
adding intra-atomic terms suppresses the gauge invariance. To circumvent this problem B.A. Foreman 
\cite{Foreman02} used group theory arguments to construct the basis in which intra-atomic matrix 
elements are present and the lattice gauge theory to define the interaction of electromagnetic 
fields with electrons in crystals. 

The effect of orbital overlapping on electronic structure has been studied 
for simple systems \cite{McKinnon95,Aldao94}.  In this paper we investigate the optical 
matrix elements in the presence of non-orthogonal 
(overlapping) orbitals. As far as we know, no study has been done in this 
direction. We show that intra-atomic contributions of the coordinate operator are induced simply by the 
orthogonalization process. The orthogonalization process induces terms 
equivalent with more distant interactions, such that it gives formal justification for the nearest-neighbor spin-orbit 
interaction introduced in Ref. \onlinecite{Boykin-so} for TB model with spin-orbit interaction\cite{Chadi77}. 
The analysis of simple systems shows that the non-orthogonal 
orbitals play an important role on optical matrix elements. We re-analyze the example of 
Pedersen \textit{et al.} \cite{Pedersen01} to show that the non-orthogonal orbitals 
improve the optical matrix elements. In the case of graphene, the overlap and TB parameterization are crucial in explaining 
the experimental data. Moreover, 
similar arguments can be employed in the \textit{ab-initio} TB-LMTO (tight binding linear 
muffin-tin orbitals) method \cite{Turek02}, leading to faster calculations of optical 
matrix elements in a parameter free theory.

\section{Tight-binding calculations and non-orthogonality}

To fix ideas we consider a localized basis $\left| {\left. {\alpha R} 
\right)} \right.$ , where \textit{$\alpha $} is the orbital type and $R$ is the center of the 
orbital (L\"{o}wdin orbitals)\cite{Lowdin50}. The crystal Hamiltonian $H$ is diagonalized 
within the Bloch sums of the localized basis 

\begin{equation}
\label{eq:1}
\left| {\left. {\alpha {\kern 1pt} k} \right)} \right. = \raise0.7ex\hbox{$1$} 
\!\mathord{\left/ {\vphantom {1 {\sqrt N 
}}}\right.\kern-\nulldelimiterspace}\!\lower0.7ex\hbox{${\sqrt N 
}$}\sum\limits_R {e^{ikR}\left| {\left. {\alpha {\kern 1pt} R} \right)} 
\right.} 
\end{equation}

\noindent
as follows

\begin{equation}
\label{eq:2}
\left| {nk} \right\rangle = \sum\limits_\alpha {c_{n\alpha } \left( k 
\right)\left| {\left. {\alpha {\kern 1pt} k} \right)} \right.} ,
\end{equation}

\noindent
with

\begin{equation}
\label{eq:3}
H\left( k \right)\left| {nk} \right\rangle = E_{nk} \left| {nk} 
\right\rangle .
\end{equation}

The kinematic momentum operator involved in optical transitions is 
defined as

\begin{equation}
\label{eq:4}
 p = \frac{m}{i\hbar }\left[ {r,H} \right].
\end{equation}

In the crystal momentum representation \cite{Blount61}, 
the kinematic momentum operator is 

\begin{equation}
\label{eq:5}
 p = \frac{m}{\hbar }\nabla _k H\left( k \right),
\end{equation}

\noindent
where $H\left( k \right)$ is the Hamiltonian in the crystal momentum 
representation. Eq.~(\ref{eq:5}) holds in a complete basis as well as in an 
incomplete basis.  However, in an incomplete basis the momentum operator \textit{p} and 
coordinate operator \textit{r} do not satisfy the canonical commutation relations leading 
to different formula for effective masses and Peierls-coupling formula 
involving the vector potential.  These issues are detailed in Refs. \onlinecite{Boykin95} and \onlinecite{Boykin01}.
The coordinate operator \textit{r} is considered to 
have the following matrix elements in the localized basis $\left| {\left. 
{\alpha R} \right)} \right.$

\begin{equation}
\label{eq:6}
\left( {\alpha 'R'} \right.\left|r\right|\left. {\alpha R} \right) 
= \left( {R\delta _{\alpha \alpha '} + d_{\alpha \alpha '} } \right){\kern 
1pt} \delta _{RR'} ,
\end{equation}

\noindent
since the overlapping of the orbitals belonging to different atoms is supposed 
to be small. Here $d_{\alpha \alpha '} $ is the intra-atomic matrix element. 
In the usual tight-binding theory the coordinate operator is diagonal
\cite{Graf95}. Therefore the intra-atomic parts are neglected \cite{Pedersen01,Cruz99} 
leading to no need of 
other fitting parameters beyond those of the Hamiltonian and to gauge 
invariance. 

Pedersen \textit{et al.} \cite{Pedersen01} pointed out that there are cases in 
which the neglect of the intra-atomic part may conduct to the underestimation of 
the momentum operator arguing that by using Eqs.~(\ref{eq:1})-~(\ref{eq:4}),

\begin{widetext}
\begin{equation}
\label{eq:8}
\begin{array}{l}
 \left\langle {nk} \right|p\left| {mk} \right\rangle = \frac{i{\kern 
1pt} m}{\hbar {\kern 1pt} }\sum\limits_{\alpha ,\alpha '} {c\ast _{n\alpha 
'} \left( k \right)\,c_{m\alpha } \left( k \right)} \;\nabla _k \left( 
{\alpha 'k} \right.\left| H \right|\left. {\alpha k} \right) + \\ 
 \quad \quad \quad \quad \;\;\quad \frac{i{\kern 1pt} m}{\hbar {\kern 1pt} 
}\left\{ {\left. {\varepsilon _{nk} - \varepsilon _{mk} } \right\} } 
\right.\sum\limits_{\alpha ,\alpha '} {c\ast _{n\alpha '} \left( k 
\right)\,c_{m\alpha } \left( k \right)\;d_{\alpha '\alpha } } \\ 
 \end{array}.
\end{equation}
\end{widetext}

The neglect of the second term in Eq.~(\ref{eq:8}) reproduces Eq.~(\ref{eq:5}). Therefore 
in the tight-binding basis, which is finite, by using Eq.~(\ref{eq:5}) one is neglecting the second term in 
Eq.~(\ref{eq:8}) or the intra-atomic part\cite{BoykinVogl02}. This shortcoming happens 
because the momentum and 
position operators do not satisfy the canonical commutation relations in a finite 
basis \cite{BoykinVogl02}. One way to add intra-atomic terms is the construction 
of B. A. Foreman \cite{Foreman02}. However, intra-atomic terms can be induced if one 
considers a non-orthogonal basis. To show this, let us have an atomic basis with 
non-zero overlapping

\begin{equation}
\label{eq:9}
\left. {\left( {\chi _{\alpha R} } \right.} \right|\left. {\chi _{\alpha 
'R'} } \right) = 1 + S_{\alpha \alpha 'RR'} = 1 + \mbox{S}.
\end{equation}

The orthogonal basis corresponding to Eq.~(\ref{eq:9}) is (the L\"{o}wdin procedure)

\begin{equation}
\label{eq:10}
\left| {\left. {\chi '} \right)} \right. = \left( {1 + \mbox{S}} \right)^{ - 
\raise0.5ex\hbox{$\scriptstyle 
1$}\kern-0.1em/\kern-0.15em\lower0.25ex\hbox{$\scriptstyle 2$}}\left| 
{\left. \chi \right)} \right.
\end{equation}

In the new orthogonal basis an operator transforms according to

\begin{equation}
\label{eq:11}
A' = \left( {1 + S} \right)^{ - \raise0.5ex\hbox{$\scriptstyle 
1$}\kern-0.1em/\kern-0.15em\lower0.25ex\hbox{$\scriptstyle 2$}}A\left( {1 + 
S} \right)^{ - \raise0.5ex\hbox{$\scriptstyle 
1$}\kern-0.1em/\kern-0.15em\lower0.25ex\hbox{$\scriptstyle 2$}}.
\end{equation}

Formally, expanding Eq.~(\ref{eq:10}) in power series of $S$ we rewrite Eq.~(\ref{eq:11}) as

\begin{equation}
\label{eq:12}
A' = A - \frac{1}{2}\left( {S{\kern 1pt} A + A{\kern 1pt} S} \right) + 
\frac{3}{8}\left( {ASS + SSA} \right) + \frac{1}{4}SAS\ldots .
\end{equation}

The inverse transform of Eq.~(\ref{eq:11}) has the following expansion

\begin{equation}
\label{eq:13}
A = A' + \frac{1}{2}\left( {S{\kern 1pt} A' + A'{\kern 1pt} S} \right) - 
\frac{1}{8}\left( {A'SS + SSA'} \right) + \frac{1}{4}SA'S\ldots 
\end{equation}

\begin{figure*}
\includegraphics{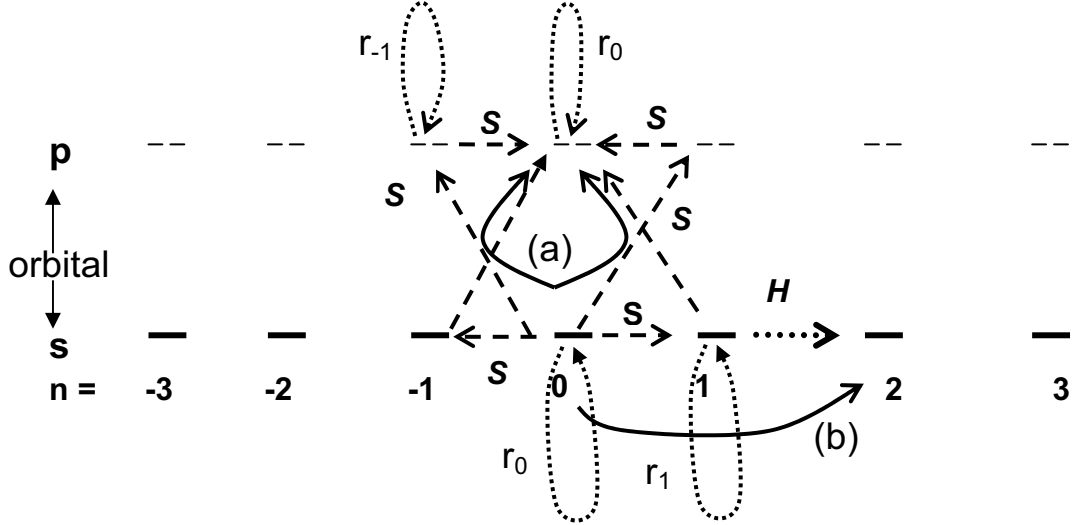}
\caption{\label{fig:1} (a) Schematic representation of the intra-atomic matrix elements of the coordinate operator 
 induced by the orbital overlap; the generic model has two orbitals per site, s-like (-) orbitals and p-like (- -) orbitals . 
We illustrate the matrix elements \textit{r} of the coordinate operator in the orthogonal basis by small dash arrows , 
the overlap matrix elements \textit{S} by dash arrows, the matrix elements \textit{d} of the coordinate operator in 
non-orthogonal basis by  full arrows, and the Hamiltonian matrix elements \textit{H} in the non-orthogonal basis by dotted arrows. (b) 
Schematic representation of the 
increase in the range of the Hamiltonian in the orthogonalized basis by using the first order approximation in Eq. (\ref{eq:11}). 
The new long-range matrix element of the Hamiltonian in the orthogonal basis connecting site 0 with site 2 for a nearest-neighbor 
Hamiltonian is shown by full arrow.}
\end{figure*}

Now suppose that in the L\"{o}wdin basis the intra-atomic matrix element 
$d_{\alpha \alpha '} $ is zero, such that the Hamiltonian fulfils the gauge 
invariance conditions. In the original non-orthogonal (atomic) basis, 
however there are intra-atomic elements. These can be easily seen if one 
applies the inverse transform Eq.~(\ref{eq:13}) (Fig.~\ref{fig:1}a). Thus in the atomic 
basis, up to the second order in $S$, the intra-atomic matrix element is 

\begin{widetext}
\begin{equation}
\label{eq:14}
d_{R\alpha R\alpha '} = \frac{1}{8}\sum\limits_{R''\alpha ''} {\left( { - 
r_R S_{R\alpha ',R''\alpha ''} {\kern 1pt} S_{R''\alpha '',R\alpha } - 
S_{R\alpha ',R''} S_{R''\alpha '',R\alpha } {\kern 1pt} r_R + 2S_{R\alpha 
',R''\alpha ''} r_{R''} S_{R''\alpha '',R\alpha } } \right)} .
\end{equation}
\end{widetext}

Eq.~(\ref{eq:14}) shows us that in the atomic basis the intra-atomic matrix elements of 
coordinate operator are non-zero. Hence, considering the overlap, 
intra-atomic optical transitions can be incorporated. Although the intra-atomic corrections are 
second order in the overlap $S$, the overall corrections to the optical matrix 
elements are first order in $S$. In the same time the 
range of the Hamiltonian has been increased by applying the transformation given by Eq.~(\ref{eq:12}) to 
the Hamiltonian matrix (Fig.~\ref{fig:1}b). 
This result suggests that although a nearest-neighbor Hamiltonian might 
give a good reproducibility of the electronic structure, it completely misses 
the intra-atomic terms of optical matrix elements.   The relationship between the 
overlap and longer ranged Hamiltonians is able to explain the nearest-neighbor 
spin-orbit interaction introduced in Ref. \onlinecite{Boykin-so} in order to reproduce the 
Dresselhaus terms in zinc blend structures. Thus, the spin orbit contribution 
to the optical matrix 
elements, $\nabla _k H_{SO} \left( k \right)$, is non-zero. In the same time 
the initial prescription given by Chadi \cite{Chadi77} is preserved. The overlap and long-range Hamiltonians 
are also closely interrelated  in quantum wire transport. Using nearest-neighbor Hamiltonians, 
the overlap is crucial in explaining anti-resonances in quantum wires \cite{Emberly99}. 
However the same anti-resonances are reproduced with a Hamiltonian in which the effect of overlapping 
has been transferred to the second nearest-neighbor hoping elements\cite{Rittenhouse05}.

In a recent paper \cite{Turek02} it is shown that a piece-wise constant coordinate operator (and therefore 
diagonal) in ab-initio TB-LMTO (tight binding linear muffin-tin orbitals) methods is 
analogous to the coordinate operator in semi-empirical methods. The most 
localized representation (TB representation), where the Hamiltonian is 
short-ranged, is not the best for calculations although it is advantageous for 
numerical treatments. On the contrary, the coordinate operator was considered 
diagonal in the (nearly) orthogonal representation (with a long-ranged 
Hamiltonian) and used in transport calculations. The results were in very 
good agreement with the experimental values and with the results with the exact 
evaluation of the coordinate operator. Thus non-orthogonality plays an important role not only in 
empirical models but also in first-principles methods. If one assumes that the coordinate operator is 
piecewise constant and that the assumption is good enough, the calculations 
of the optical matrix elements can be obtained faster from electron band 
calculations. From Eq.~(\ref{eq:8}), the \textit{k}-derivative of the Hamiltonian is calculated by 
fast Fourier transformations. Thus it is more computationally efficient than 
the usual scheme presented in Ref. \onlinecite{Lambrecht00}. However, the applicability of a
piecewise constant coordinate to optical properties of various physical systems remains 
to be investigated.

\section{Optical matrix elements in simple tight-binding models with overlap}

In the following we analyze the one-dimensional monoatomic crystal with two 
orbitals per atom, the one-dimensional diatomic crystal with one orbital per 
atom, and the two-dimensional graphene. 

\paragraph{monoatomic chain with two orbitals per atom.}
 Schematic representation of a monoatomic chain with two orbitals per atom is given in Fig.~\ref{fig:1}. 
In a Bloch basis constructed from the 
overlapping orbitals, the nearest-neighbor tight-binding Hamiltonian for a 
monoatomic chain with two orbitals per site is a 2x2 matrix

\begin{equation}
\label{eq:15}
H\left( k \right) = \left( {{\begin{array}{*{20}c}
 {E_S + 2{\kern 1pt} {\kern 1pt} V_{SS} \cos {\kern 1pt} {\kern 1pt} \left( 
{kL} \right)} \hfill & {2{\kern 1pt} {\kern 1pt} i{\kern 1pt} {\kern 1pt} 
{\kern 1pt} V_{SP} \sin {\kern 1pt} {\kern 1pt} \left( {kL} \right)} \hfill 
\\
 { - 2{\kern 1pt} {\kern 1pt} {\kern 1pt} i{\kern 1pt} {\kern 1pt} V_{SP} 
\sin {\kern 1pt} {\kern 1pt} \left( {kL} \right)} \hfill & {E_P + 2\,V_{PP} 
\cos {\kern 1pt} {\kern 1pt} \left( {kL} \right)} \hfill \\
\end{array} }} \right),
\end{equation}

\noindent
where $ E_{S}$ and $E_{P}$ are the energies of $s$-like and $p$-like orbitals, 
respectively, $V_{SS}$ and $V_{PP}$ are the coupling of two nearest neighbor 
$s$-like and $p$-like orbitals, respectively, and $V_{SP}$ is the coupling of a 
$s$-like orbital with the nearest neighbor $p$-like orbital. $L$ is the length of the 
unit cell and $k$ is the wave vector. The overlap matrix has a similar form

\begin{equation}
\label{eq:16}
S\left( k \right) = \left( {{\begin{array}{*{20}c}
 {1 + 2S_{SS} \cos {\kern 1pt} {\kern 1pt} \left( {kL} \right)} \hfill & 
{2{\kern 1pt} i{\kern 1pt} S_{SP} \sin {\kern 1pt} {\kern 1pt} \left( {kL} 
\right)} \hfill \\
 { - 2{\kern 1pt} i{\kern 1pt} S_{SP} \sin {\kern 1pt} {\kern 1pt} \left( 
{kL} \right)} \hfill & {1 + 2S_{PP} \cos {\kern 1pt} {\kern 1pt} \left( {kL} 
\right)} \hfill \\
\end{array} }} \right).
\end{equation}

\begin{figure}
\includegraphics{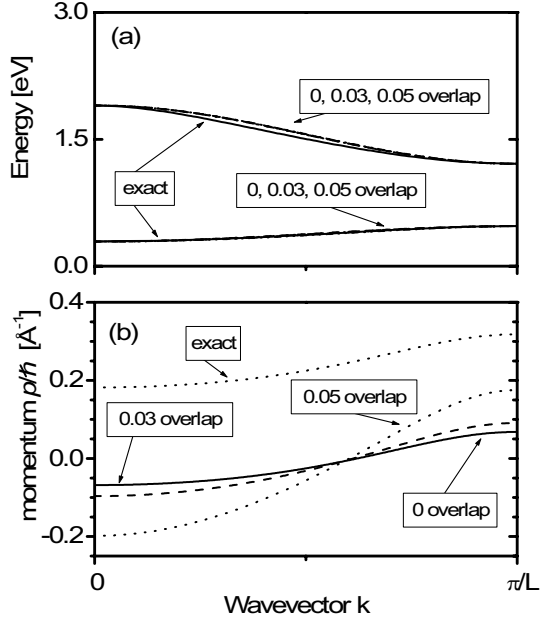}
\caption{\label{fig:2} (a) Band structure of the exact Kroning-Penney model 
and its approximations with a tight-binding model with overlap (see the text). 
Notice that following the same fitting procedure, the approximate bands are indistinguishable.
 (b) Momentum matrix elements of the exact Kroning-Penney model and its approximations with 
a tight-binding model with overlap. Notice the large variation of the momentum matrix elements 
with respect to the overlap while the bands are almost identical.}
\end{figure}

\noindent
In the orthogonal basis constructed according to Eq.(~\ref{eq:10}), the 
Hamiltonian matrix $\tilde {H}$ is given by Eq.~(\ref{eq:11}). The electronic 
bands are given by solving the eigenvalue problem 
$\tilde {H}\left( k \right)\left| {\left. {nk} \right\rangle } \right. = 
E_{nk} \left| {nk} \right\rangle $ and the interband matrix element of the 
kinematic momentum operator is $p\left( k \right) = \frac{m}{\hbar 
}\left\langle {1k} \right|\tilde {H}'\,\left( k \right)\left| {2k} 
\right\rangle $ , with $\tilde {H}'$ the derivative of $\tilde {H}$ 
with respect to $k$. We apply the above model to approximate the lowest two 
bands of the one-dimensional Kronig-Penny model. The Kronig-Penney model is 
a set of quantum wells of width $a$ separated by barriers of height $V_{0}$ and width 
$b$. The case is investigated by Pedersen \textit{et al.} \cite{Pedersen01} 
to suggest the need for intra-atomic contributions to optical transitions. 
We consider their strong-coupling case with $a = $8 {\AA}, $b = $1 {\AA}, 
and $V_{0}$ = 5 eV. The first state in the quantum well is an $s$-like state, while second state is a 
$p$-like state. Accordingly, in the tight-binding counterpart of the 
Kronig-Penney model, the overlap matrix elements $S_{SP}$ and $S_{PP}$ have to be 
negative. We adopt the same procedure\cite{Pedersen01} for fitting the energy bands 
of the Kronig-Penney model. The absolute values of the overlap matrix 
elements are chosen to be the same for $S_{SS}$, $S_{SP}$, and $S_{PP}$. The 
results are shown in Fig.~\ref{fig:2} for an overlap of 0, 0.03 and 0.05 in comparison 
with the exact results of the Kronig-Penney model. While the energy bands 
are indistinguishable for tight-binding counterparts and agree well with the 
exact values, the interband momentum matrix elements vary and move toward 
exact values of the Kronig-Penney model. Because the absorption spectra are 
determined by the square modulus of the momentum matrix elements the above 
result is quite remarkable in the following sense as we explain below. 
Although we considered 
the strong coupling case (thin barriers), the coupling between $s$-like and 
$p$-like states is weak (the matrix element $V_{SP}$ is an order of magnitude 
smaller than the other matrix elements) such that the electron bands 
have almost either $s$- or $p$-like character over the entire Brillouin zone. 
Therefore, $V_{SP}$ determines the magnitude of the interband momentum matrix 
element. In the same time the validity of $p\left( k \right) = 
\frac{m}{\hbar }\left\langle {1k} \right|\tilde {H}'\,\left( k 
\right)\left| {2k} \right\rangle $ is appropriate for strong 
inter-atomic coupling, such that the nearest-neighbor tight-binding model with orthogonal orbitals 
is inappropriate 
to calculate optical properties for the above model. Finally we want to mention that in one-dimensional 
crystals with inversion symmetry the coordinate operator is diagonal in the 
basis generated by the Wannier functions \cite{Kivelson83}. Hence, the ``closer'' 
to the Wannier functions are the L\"{o}wdin orbitals, the better reproduced 
are the momentum matrix elements.

\begin{figure}
\includegraphics{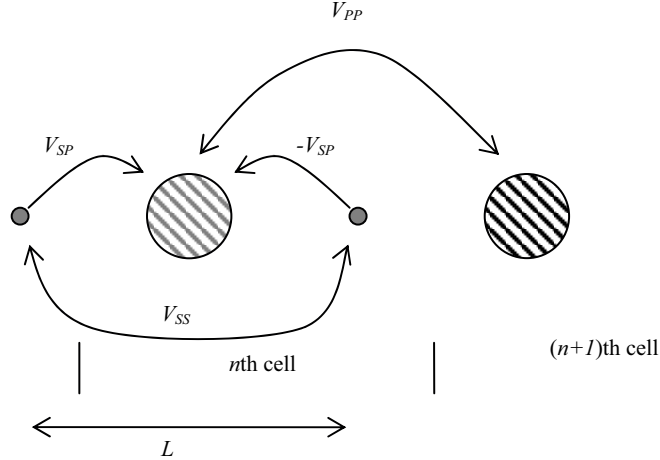}
\caption{\label{fig:3} Schematic representation of the diatomic linear chain with one 
 orbital per atom and lattice constant \textit{L} . The first type of atoms are depicted 
as small and full circles and the second type as stripped circles. The interactions between 
atoms are shown by arrows.}
\end{figure}

\paragraph{One-dimensional diatomic crystal with one orbital per atom.} 
The chain is 
represented by $s$-like orbitals at positions \textit{nL} and $p$-like orbitals 
at $n{\kern 1pt} L + L \mathord{\left/ {\vphantom {L 2}} \right. 
\kern-\nulldelimiterspace} 2$, where $L$ is length of unit cell and $n$ is 
integer. The interaction up to the second-nearest neighbor is illustrated in 
Fig.~\ref{fig:3}. The corresponding Hamiltonian matrix is

\begin{equation}
\label{eq:17}
H = \left( {{\begin{array}{*{20}c}
 {E_S + 2{\kern 1pt} {\kern 1pt} V_{SS} \cos {\kern 1pt} {\kern 1pt} \left( 
{kL} \right)} \hfill & {2{\kern 1pt} {\kern 1pt} i{\kern 1pt} {\kern 1pt} 
{\kern 1pt} V_{SP} \sin {\kern 1pt} {\kern 1pt} \left( {{kL} \mathord{\left/ 
{\vphantom {{kL} 2}} \right. \kern-\nulldelimiterspace} 2} \right)} \hfill 
\\
 { - 2{\kern 1pt} {\kern 1pt} {\kern 1pt} i{\kern 1pt} {\kern 1pt} V_{SP} 
\sin {\kern 1pt} {\kern 1pt} \left( {{kL} \mathord{\left/ {\vphantom {{kL} 
2}} \right. \kern-\nulldelimiterspace} 2} \right)} \hfill & {E_P + 2\,V_{PP} 
\cos {\kern 1pt} {\kern 1pt} \left( {kL} \right)} \hfill \\
\end{array} }} \right).
\end{equation}

\begin{figure}
\includegraphics{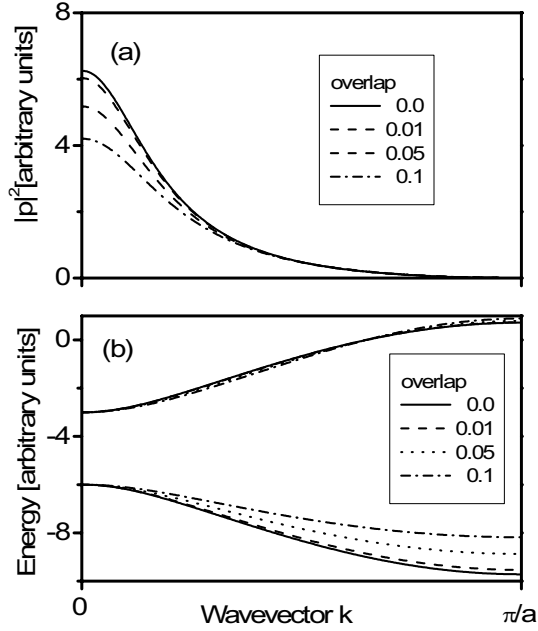}
\caption{\label{fig:4} (a) Optical matrix elements of the tight-binding model with overlap for a 
diatomic linear chain with one orbital per atom. (b) Energy bands of the tight-binding model with 
overlap for a diatomic linear chain with one orbital per atom. Full line is for 0 overlap, dash line 
for a 0.01 overlap, dot line for a 0.05 overlap, and dash-dot line for a 0.1 overlap. We choose 
arbitrary units because the system is rather generic.}
\end{figure}

Similar form holds for the overlap matrix. This can be an approximate 
model for superlattices of type II, such as InAs-GaSB. In the InAs-GaSB 
superlattice the central feature 
is that the top of the GaSb valence band lies higher in energy than the 
bottom of the InAs conduction band, such that the electron and hole wave 
functions are overlapping. The electron/hole wave function is modeled by 
$s$-like/$p$-like orbitals. Keeping only the nearest neighbor interaction and overlap, 
the effect of overlapping is to decrease the momentum matrix elements as it 
is shown in Fig.~\ref{fig:4}. This simple result might help in explaining the increase 
of the photoluminescence intensity with the reduction of the electron-hole 
wave function overlap\cite{Ongstad01}, which is not explained by the empirical pseudo-potential 
calculations used to for this purpose\cite{Dente99}. The  empirical pseudo-potential method\cite{Dente99} used is 
non-atomistic, i.e. in their approach the Hamiltonian of the InAs/GaSb superlattice is constructed from the 
potential form factors of the InAs and GaSb bulk constituents. The potentials of the two bulk constituents 
are matched continuously at the interfaces such that there are no In-Sb or Ga-As bonds at the interface as there must be. 
As pointed out in Ref.~\onlinecite{Magri03} an 
atomistic description is desired to take into account charge redistribution, segregation, and interdiffusion at the 
interface between InAs and GaSb. In contrast to Ref. ~\onlinecite{Dente99}, Magri and Zunger~\cite{Magri03} solve the 
single-particle Schr\"{o}dinger equation for each atom in the structure making their method atomistic. In this sense, 
TB models preserve the atomistic description of interfaces.

\begin{figure*}
\includegraphics{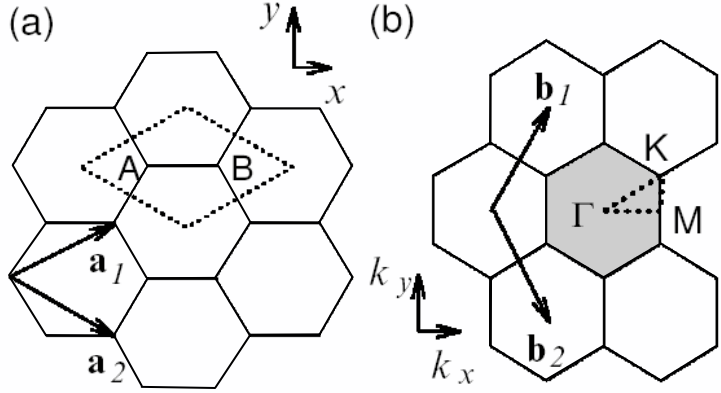}
\caption{\label{fig:5} (a) The unit cell of two-dimensional graphite is shown as 
the dotted rhombus. (b) The Brillouin zone of two-dimensional graphite is shown as 
the shaded hexagon. \textbf{a}$_{i} $, and \textbf{b}$_{i} $, (i =1;2) are 
basis vectors and reciprocal lattice vectors, respectively. Energy dispersion relations and 
optical matrix elements are calculated along the perimeter of the dotted triangle connecting the high symmetry 
points, \textit{$\Gamma$}, \textit{$K$}, and \textit{$M$}.}
\end{figure*}

\paragraph{Graphene.} 
Recently, graphene as a two-dimensional sheet of graphite has been widely 
studied in the context of carbon nanotubes \cite{Saito98}. Graphite consists of a stack of 
graphene sheets , piled up and weakly interacting one with each other. Graphene has a 
hexagonal structure with two atoms in the unit cell (Fig.~\ref{fig:5}) and very strong 
\textit{sp}$^{2}$ bonds, causing a threefold coordinated planar structure. The 
remaining $p_{z}$ orbitals are perpendicular to the plane, forming \textit{$\pi $} (bonding) 
and \textit{$\pi $*} (antibonding) states. The overlap of \textit{$\pi $} electrons with the intra-plane 
\textit{sp}$^{2}$ orbitals is small and \textit{$\pi $} and \textit{$\pi $*} electronic states dominate the physical 
properties at low energy, around Fermi level. From Fig.~\ref{fig:5} we easily deduce 
the nearest-neighbor tight-binding Hamiltonian and overlap matrix for \textit{$\pi $} and 
\textit{$\pi $*} states as

\begin{equation}
\label{eq:18}
H\left( k \right) = \left( {{\begin{array}{*{20}c}
 {E_P } \hfill & {\gamma _0 f\left( k \right)} \hfill \\
 {\gamma _0 f^\ast \left( k \right)} \hfill & {E_P } \hfill \\
\end{array} }} \right)
\end{equation}

\noindent
and

\begin{equation}
\label{eq:19}
S\left( k \right) = \left( {{\begin{array}{*{20}c}
 1 \hfill & {s_0 f\left( k \right)} \hfill \\
 {s_0 f^\ast \left( k \right)} \hfill & 1 \hfill \\
\end{array} }} \right),
\end{equation}

\noindent
with $f\left( k \right) = e^{{ik_y a} \mathord{\left/ {\vphantom {{ik_y a} 
{\sqrt 3 }}} \right. \kern-\nulldelimiterspace} {\sqrt 3 }} + 2e^{{ - ik_y 
a} \mathord{\left/ {\vphantom {{ - ik_y a} {2\sqrt 3 }}} \right. 
\kern-\nulldelimiterspace} {2\sqrt 3 }}\mbox{cos}\left( {\frac{k_x a}{2}} 
\right)$, \textit{$\gamma $}$_{0}$ is the nearest neighbor transfer integral, $E_{p}$ is the energy 
of \textit{$\pi $} orbitals, $s_{0}$ is the nearest neighbor overlap integral, a(=0.246 nm) is 
the lattice constant of graphite, and \textit{k} is the two 
dimensional wave vector. Experimental data or first principles calculations 
put \textit{$\gamma $}$_{0}$ between 2.5 and 3 eV, $E_{p}$ = 0 eV , and $s_{0}$ is found to be 
below 0.1 \cite{Reich02}. Due to their similar form, Hamiltonian matrix and overlap 
matrix have the same eigenvectors $\left| {u^\pm } \right\rangle = \left( 
{\raise0.7ex\hbox{$1$} \!\mathord{\left/ {\vphantom {1 
2}}\right.\kern-\nulldelimiterspace}\!\lower0.7ex\hbox{$2$},\raise0.7ex\hbox{${ 
\mp e^{ - i\varphi }}$} \!\mathord{\left/ {\vphantom {{ \mp e^{ - i\varphi 
}} 2}}\right.\kern-\nulldelimiterspace}\!\lower0.7ex\hbox{$2$}} \right)$, 
with \textit{$\varphi $} defined as $f\left( k \right) = \left| {f\left( k \right)} 
\right|\,e^{i\varphi \left( k \right)} = w\left( k \right)\,e^{i\varphi 
\left( k \right)}$. This yields the electronic eigenvalues 

\begin{equation}
\label{eq:20}
E^\pm \left( k \right) = \frac{E_p \mp \gamma _0 w\left( k \right)}{1 \mp 
s_0 w\left( k \right)}.
\end{equation}

We note that the overlap makes the energy bands asymmetric with respect to 
the Fermi level and has large influence on bands. The full form of the 
Hamiltonian with overlap, $\tilde {H}$, is

\begin{equation}
\label{eq:20a}
\tilde {H}\left( k \right) = \frac{1}{1 - s_0 ^2w^2\left( k \right)}\left( 
{{\begin{array}{*{20}c}
 {E_P - s_0 {\kern 1pt} \gamma _0 w^2\left( k \right)} \hfill & {\left( 
{\gamma _0 - s_0 E_p } \right)f\left( k \right)} \hfill \\
 {\left( {\gamma _0 - s_0 {\kern 1pt} E_P } \right)f^\ast \left( k \right)} 
\hfill & {E_P - s_0 {\kern 1pt} \gamma _0 w^2\left( k \right)} \hfill \\
\end{array} }} \right)
\end{equation}

Since $s_{0}$ is less than 0.1 we can safely discard the prefactor in Eq.~(\ref{eq:20a}). The 
diagonal part of the Hamiltonian matrix is proportional to the unit 
matrix in both cases, with or without overlap, and it does not contribute to 
the interband momentum matrix element. Therefore, one can easily calculate 
the intraband momentum matrix element in a compact form as

\begin{equation}
\label{eq:21}
p = \frac{m}{\hbar }\left\langle {u^ + } \right|\nabla \tilde {H}\left( k 
\right)\left| {u^ - } \right\rangle = \frac{i{\kern 1pt} m}{\hbar }\left( 
{\gamma _0 - E_p s_0 } \right)\frac{w\left( k \right)}{1 - s_0 ^2w^2\left( k 
\right)}\nabla \varphi \left( k \right)
\end{equation}

\begin{figure}
\includegraphics{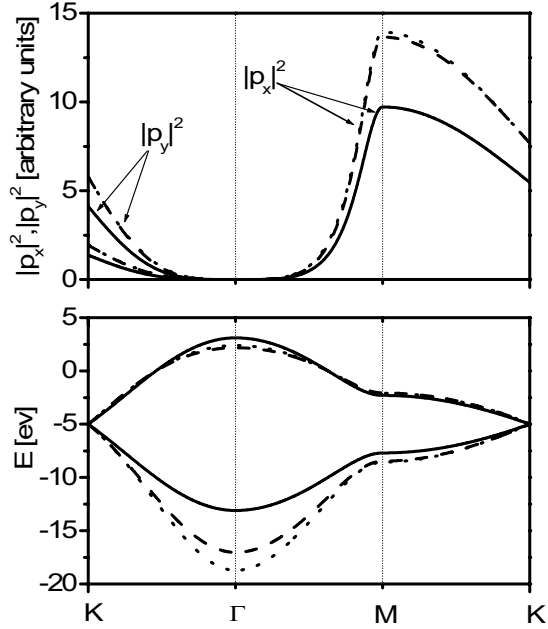}
\caption{\label{fig:6} (a) Optical matrix elements of the tight-binding model with overlap for the  
two-dimensional graphene. Since we are interested in the relative change of the momentum matrix 
elements with respect to overlap, arbitrary units are used. 
(b) Energy bands of the tight-binding model with 
overlap for the two-dimensional graphene. The values used are the following: 
\textit{$\gamma $}$_{0}$ = 2.7 eV, $E_{p}$ = -5 eV, and $s_{0}$ =0.1.}
\end{figure}

Eq.~(\ref{eq:21}) shows us that for the most used parametrization ($E_{p}$ = 0 eV) ,
the overlap does not play any role on the inter-band momentum since $s_{0}$ is 
less than 0.1 and we can safely discard the second order term is $s_{0}$. However, in order to fit 
the experimental dielectric function with the nearest-neighbor model, orbital overlapping 
is invoked in Ref.~\onlinecite{Pedersen03}.  It was found that \textit{$\gamma $}$_{0}$ = 2.7 eV 
and $E_{p}$ = -5 eV by assuming $s_{0}$ =0.1. With this  parameterization the numerical results are 
shown in Fig.~\ref{fig:6} for the electronic bands and momentum matrix elements.  The momentum 
matrix elements for the case with overlap are practically the same as those 
of the first order approximation Hamiltonian (Eq.~(\ref{eq:12})). In the same time, the electronic bands 
generated by the first order Hamiltonian are different from those of the 
full Hamiltonian with overlap.

\section{Conclusions}

We investigated the influence of the non-orthogonal orbitals on optical 
matrix elements in tight-binding models. A diagonal coordinate operator in 
the orthogonalized basis not only ensures the gauge invariance but also 
induces intra-atomic contributions to the coordinate operator in the original 
(atomlike and non-orthogonal) basis. Moreover, the Hamiltonian matrix in the 
orthogonal basis is longer ranged than the Hamiltonian matrix in the initial 
non-orthogonal basis. As a consequence, one can justify the nearest-neighbor 
interaction of the spin-orbit coupling\cite{Boykin-so}. It enables to describe the Dresselhaus term,
which is not considered in the usual treatment of the spin-orbit coupling\cite{Chadi77}.

Simple models are analyzed. The first model studied was the monoatomic linear chain with 
two orbitals per site as 
an approximation to the Kronig-Penney model. The model was also used in Ref.
\onlinecite{Pedersen01} to show the role played by the intra-atomic matrix elements of the
momentum operator. We found that, although the tight-binding model with 
overlap exhibits almost the same energy bands as the one with orthogonal orbitals, 
the optical matrix elements 
are closer to the exact matrix elements of Kronig-Penney model. The second model 
studied was the biatomic linear chain with one orbital per site. This case 
showed that optical matrix elements decrease with overlap increasing. We also 
analyzed the optical matrix elements of the tight-binding model for two-dimensional graphite at low energies 
(between \textit{$\pi $} and \textit{$\pi $*} electronic states). Optical matrix 
elements remain unchanged 
with respect to the overlap when the usual parametrization $E_{P}$= 0 eV is 
adopted, while the bands change drastically. However, non-vanishing orbital overlapping 
and $E_{P}$= -5.0 eV are needed for better agreement with experimental data \cite{Pedersen03}. 

In complete analogy with the above arguments, one can use a piecewise 
constant coordinate operator in the orthogonal representation of tight-binding 
linear muffin-tin orbitals methods\cite{Turek02} to calculate optical spectra from 
ab-initio. The procedure will be faster because it will enable to calculate optical 
spectra directly from energy band calculations by employing the fast Fourier transformation and
without directly evaluating the momentum operator.

\begin{acknowledgments}
The author wishes to acknowledge the support in part by the Office of Naval Research.
\end{acknowledgments}


\end{document}